\documentclass[12pt,aps,preprint,nofootinbib]{revtex4}
\makeatletter
\def\ps@copyright{}
\makeatother

\usepackage[a4paper, height=220mm, width=155mm]{geometry}
\usepackage{graphicx}
\usepackage{epsfig}
\usepackage{amsmath}

\newcommand{\eVq}{\ensuremath{\text{eV}^2}}
\newcommand{\Dmq}{\Delta m^2}
\newcommand{\Sol}{\text{sol}}
\newcommand{\Atm}{\text{atm}}

\begin{document}

\title{The $\gamma = 100$ $\beta$-Beam revisited} 

\author{Enrique Fernandez-Martinez}
\affiliation{Max-Planck-Institut f\"ur Physik (Werner-Heisenberg-Institut), F\"ohringer Ring 6, 80805 M\"unchen, Germany}

\begin{abstract}
We study the performance of $\gamma =100$ $\beta$-Beam setups based on the decays of $^8$B and $^8$Li as a function of the achievable production fluxes and compare them with the ``standard'' setups based on $^{18}$Ne and $^{6}$He decays. For the standard setup we also investigate the physics potential for reduced $^{18}$Ne fluxes, since it seems experimentally challenging to achieve the baseline numbers. We find that, contrary to the standard setup, setups based in $^8$B and $^8$Li can be sensitive to the mass hierarchy down to values of $\sin^2 2 \theta_{13} \sim 10^{-2}$ by themselves, due to the stronger matter effects granted by the higher energy of the neutrinos. On the other hand, the longer baseline required for neutrinos to oscillate at these higher energies reduces the statistics at the detector and fluxes around three times larger are required to reach the sensitivity to $\theta_{13}$ and CP violation for the smallest values of $\sin^2 2 \theta_{13}$ that the standard setup provides. We also studied the required suppression factor of the atmospheric background. In all the setups studied, we found that a suppression factor of $10^{-4}$ is equivalent to removing completely the atmospheric background. Suppression factors of order $10^{-3}$ do not imply a significant loss in sensitivity while suppressions of $10^{-2}$ still allow good CP discovery potential if $\sin^2 2 \theta_{13} > 10^{-2}$ and the flux is not too low. 
\end{abstract}

\preprint{MPP-2009-178}
\preprint{EURONU-WP6-09-13}

\maketitle

\section{Introduction}
Neutrino physics is one of the few evidences we have for physics beyond the Standard Model of particle physics. The results from solar~\cite{Cleveland:1998nv,Abdurashitov:1999zd,Hampel:1998xg,Fukuda:2001nj,Ahmad:2001an,Ahmed:2003kj},
atmospheric~\cite{Fukuda:1998mi,Ambrosio:2001je},
reactor~\cite{Apollonio:1999ae,Apollonio:2002gd,Boehm:2001ik,Eguchi:2002dm} 
and accelerator~\cite{Ahn:2002up,Aliu:2004sq,Michael:2006rx} 
neutrino experiments show that neutrinos change flavour in flight. The extension of the Standard Model with neutrino masses and three-family mixing that would drive neutrino oscillations is the simplest and best explanation that accommodates all data. The experimental results point to two very different mass-squared
differences, $\Dmq_\Sol \approx 7.7 \times 10^{-5}~\eVq$ and
$|\Dmq_\Atm| \approx 2.4 \times 10^{-3}~\eVq$. On the other hand, only
two out of the four parameters of the three-family leptonic mixing
matrix are known: $\theta_{12} \approx 34^\circ$ and $\theta_{23} \approx 42^\circ$~\cite{GonzalezGarcia:2007ib}.  
The other two parameters, $\theta_{13}$ and $\delta$, are still unknown: for the mixing angle
$\theta_{13}$ direct searches at reactors~\cite{Apollonio:1999ae,Apollonio:2002gd,Boehm:2001ik} and three-family global analysis of
the experimental data give the upper bound $\sin^2 2 \theta_{13} \leq 0.074$ at $1\sigma$, whereas for the leptonic CP-violating phase $\delta$ we
have no information whatsoever. Furthermore, the ordering of the neutrino mass eigenstates, 
i.e. the sign of the atmospheric mass difference $\Dmq_\Atm$, also remains unknown since atmospheric neutrino oscillations are only sensitive to its absolute value. 

Recently, a preference for $\sin^{2}\theta_{13}>0$ has been found when fitting solar neutrino data from the third phase of the Sudbury Neutrino Observatory (SNO-III) and recent data from KamLAND~\cite{Fogli:2008jx}. The first results from the MINOS experiment, studying the appearance channel $\nu_{\mu}\rightarrow\nu_{e}$, also show a preference for non-zero values of $\theta_{13}$. A combination of all these data provides a best fit of $\sin^{2}\theta_{13} \sim 0.01$, depending on the solar model assumed, and disfavor $\theta_{13}=0$ at slightly more than $1 \sigma$ ~\cite{GonzalezGarcia:2010er}. This hint for non-zero $\theta_{13}$  will be probed by the forthcoming generation of accelerator \cite{t2k,nova} and reactor \cite{chooz2,dayabay} experiments, However, even if the hint for large $\theta_{13}$ is confirmed, these experiments lack the power to probe the remaining unknown neutrino oscillation parameters,
such as the existence of leptonic CP violation encoded in the phase $\delta$ or the ordering of neutrino masses \cite{Huber:2009cw}. A new generation of neutrino oscillation experiments is therefore needed for this task or to explore even smaller values of $\theta_{13}$ if the present hint is not confirmed.

The three types of facilities proposed for the next generation of neutrino oscillation experiments are characterized by different production mechanisms of the neutrino beam. 
Super-Beams would constitute an upgrade of conventional $\nu_{\mu}$ beams from pion decay with MW proton drivers. T2K~\cite{t2k} and NO$\nu$A~\cite{nova} are often considered to be the first generation of Super-Beam experiments. A more ambitious facility, for a next to next generation upgrade, is the Neutrino Factory, which would produce very intense $\nu_\mu$ and $\bar{\nu}_e$ beams from muon decays accelerated to energies of $\sim 25$ GeV~\cite{Geer:1997iz,De Rujula:1998hd,Bandyopadhyay:2007kx}. As an intermediate step, lower energy versions of the Neutrino Factory with energies $\sim 4$ GeV have also been proposed~\cite{Geer:2007kn,Bross:2007ts,Tang:2009wp,Bross:2009gk}. Finally, $\beta$-Beams involve the production of $\beta$-unstable ions, accelerating them to some reference
energy, and allowing them to decay in the straight section of a storage ring, resulting in a very intense and pure $\nu_e$ or $\bar \nu_e$ beams~\cite{Zucchelli:sa}. The $\nu_e \to \nu_\mu$ ``golden channel'', or its T-conjugate channel, can be probed at Super-Beams, $\beta$-Beams and Neutrino Factories and has been identified as the most sensitive to all the unknown parameters \cite{Cervera:2000kp}. However, strong correlations between them \cite{Burguet-Castell:2001ez,Minakata:2001qm,Fogli:1996pv,Barger:2001yr} make 
the simultaneous measurement of all the parameters extremely difficult. 

In the original $\beta$-Beam proposal, $^{18}$Ne ($^{6}$He) ions are accelerated to $\gamma \sim 100$ at the CERN SPS and stored so that $\nu_e$ ($\bar{\nu}_e$) beams are produced and the golden channel oscillation is searched for at a Mton class water Cerenkov detector located at $L=130$ km at the Frejus site, detailed analyses of the physics performance of this setup can be found in Refs. \cite{Mezzetto:2003ub,Mezzetto:2004gs,Donini:2004hu,Donini:2004iv,Donini:2005rn,Huber:2005jk,Campagne:2006yx}. Numerous modifications of this basic setup have been studied \cite{Agarwalla:2005we,Donini:2006tt,Volpe:2006in,Agarwalla:2006vf,Donini:2007qt,Jansson:2007nm,Agarwalla:2007ai,Coloma:2007nn,Meloni:2008it,Agarwalla:2008gf,Agarwalla:2008ti,Winter:2008cn,Winter:2008dj,Choubey:2009ks}, most of them being different combinations of two basic ingredients: the possibility of accelerating the ions to higher $\gamma$ factors \cite{Burguet-Castell:2003vv,Burguet-Castell:2005pa}, thus increasing the flux and the statistics at the detector, and the possibility of considering the decay of different ions to produce the neutrino beam. In particular $^8$B and $^8$Li
have been proposed as alternatives to $^{18}$Ne and $^{6}$He respectively \cite{Rubbia:2006pi,Donini:2006dx,Rubbia:2006zv}.

Here we will focus on the $\gamma = 100$ $\beta$-Beam option and investigate its physics potential. The $\gamma \sim 100$ limitation on the boost factor of the ions stems from the requirement that the CERN SPS can already be exploited to access these energies and thus leverage existing infrastructure in the same spirit as the original proposal. Moreover, the production of the required isotopes could also profit from the EURISOL nuclear physics programme that could also be based at CERN~\cite{eurisol}.
In Section \ref{baselines} we compare the different possible choices of ions and baselines. In Section \ref{atmo} we take into account the atmospheric background expected at the detector and study the suppression factor of this background required through the bunching of the beam to attain the best sensitivities in the different setups. Finally in Section \ref{concl} we summarize our conclusions.

\section{Choice of ions and baselines}
\label{baselines}

\begin{table}
\begin{center}
\begin{tabular}{|c|c|c|c|c|} \hline \hline
   Element  & $A/Z$ & $T_{1/2}$ (s) & $Q_\beta$ eff (MeV) & Decay Fraction \\ 
\hline
  $^{18}$Ne &   1.8 &     1.67      &        3.41         &      92.1\%    \\
            &       &               &        2.37         &       7.7\%    \\
            &       &               &        1.71         &       0.2\%    \\ 
  $^{8}$B   &   1.6 &     0.77      &       13.92         &       100\%    \\
\hline
 $^{6}$He   &   3.0 &     0.81      &        3.51         &       100\%    \\ 
 $^{8}$Li   &   2.7 &     0.83      &       12.96         &       100\%    \\ 
\hline
\hline
\end{tabular}
\caption{\label{tab:ions}  $A/Z$, half-life and end-point energies for three $\beta^+$-emitters ($^{18}$Ne  and $^8$B)
and two $\beta^-$-emitters ($^6$He and $^8$Li). All different $\beta$-decay channels for $^{18}$Ne are presented.}
\end{center}
\end{table}
In Tab.~\ref{tab:ions} we show the relevant parameters for the $\beta$ decay of four ions: $^{18}$Ne and $^6$He, $^8$Li and $^8$B. As can be seen, the main difference between the two sets of ions consists in their decay energy $Q_\beta \sim 3.5$ for $^{18}$Ne and $^6$He and $Q_\beta \sim 13.5$ for $^8$Li and $^8$B. Thus, the neutrino beams produced by the decay of the latter set of ions are around $3.5$ times more energetic than the ones produced by the former when accelerated to the same $\gamma$ factor. This has the obvious advantage that, for the same $\gamma$ factors, higher neutrino energies are achievable with $^8$Li and $^8$B.
On the other hand, it also means that, for the oscillation to be on peak, a baseline $3.5$ times longer is required to achieve the same $L/E$ value. This translates into a suppression of the neutrino flux at the detector, since the flux decreases with $L^{-2}$ and the suppression is not fully compensated with the enhancement of the cross section at higher energies. For this reason, $\beta$-Beams based on $^8$Li and $^8$B decays usually suffer from low statistics. This was also the reason why ions such as $^{18}$Ne and $^6$He with rather low decay energies were chosen in the original proposal.

Despite their statistical limitations, $^8$Li and $^8$B also provide an interesting choice since higher neutrino energies can be probed with the maximum $\gamma$ factor achievable at with the same booster (SPS) as envisaged in the original $\beta$-Beam proposal~\cite{Zucchelli:sa}. The higher energies accessible translate in stronger matter effects and generally higher sensitivity to the neutrino mass hierarchy through them, even allowing to reach a resonant behaviour if $\gamma=350$ is accessible \cite{Agarwalla:2006vf,Agarwalla:2007ai,Coloma:2007nn,Agarwalla:2008ti,Choubey:2009ks}.

The choice of the baseline should match the neutrino energy, since neutrino flavour change oscillates with $L/E$. Too short a baseline will not allow oscillations to develop and a baseline much larger than the one matching the first oscillation peak will reduce unnecessarily the statistics at the detector due to the beam divergence as $L^{-2}$.
We will consider two possible sites capable of housing the Mton class water Cerenkov detector proposed to observe the neutrino beam produced at CERN and that match the baseline requirements. The shorter baseline is $130$ km and matches the CERN to Frejus distance, it is suitable to observe the first oscillation peak of neutrinos from $^{18}$Ne and $^6$He decays accelerated to $\gamma =100$, this was the baseline suggested in the original $\beta$-Beam proposal. A longer baseline of $650$ km would correspond to the CERN to Canfranc distance and would roughly match the first oscillation peak of neutrinos from $^8$Li and $^8$B decays accelerated to $\gamma =100$. Oscillations of neutrinos from $^{18}$Ne and $^6$He could also be observed at this baseline at the second peak for $\gamma=100$ or at the first peak for $\gamma=350$ (achievable with a refurbished SPS~\cite{Burguet-Castell:2003vv}).

A key factor in the determination of the best ion candidates is the achievable number of decays per year for each ion. This factor is at present extremely uncertain. For 
$^{18}$Ne and $^6$He ``standard'' fluxes of $1.1 \cdot 10^{18}$ and $2.9 \cdot 10^{18}$ decays per year are usually assumed. These fluxes would grant the $\gamma =100$ $\beta$-Beam proposal enough sensitivity to compete with similar Super-Beam facilities. Preliminary studies show that this requirement should be achievable for $^6$He ions, the estimations actually yield a flux somewhat larger. In the case of $^{18}$Ne, on the other hand, the production of an intense flux is much more challenging and the present estimates yield a flux about an order of magnitude smaller. More sophisticated productions could improve the situation to a factor 5 smaller than the ``standard'' flux~\cite{ne18,Lindroos:2010zz}.

\begin{figure}[t!]
\vspace{-0.5cm}
\begin{center}
\hspace{-0.7cm} \epsfxsize7.5cm\epsffile{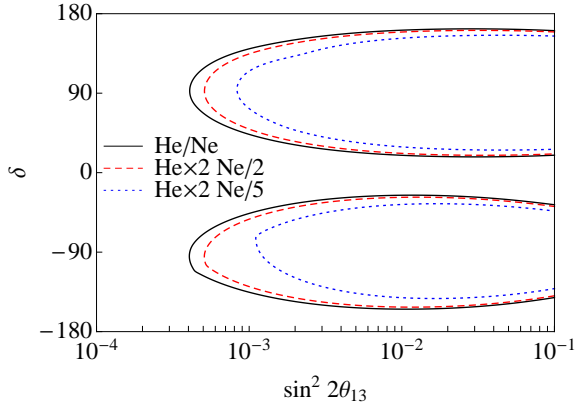} 
\caption{\label{fig:lowne}
Comparison of the 3 $\sigma$ discovery potential to leptonic CP violation of the standard $^{18}$Ne and $^6$He with different assumptions for the achievable ion fluxes.}
\end{center}
\end{figure}

In Fig.~\ref{fig:lowne} we compare the CP discovery potential, defined as the values of $\theta_{13}$ and $\delta$ that would allow to discard at $3 \sigma$ the CP conserving values $\delta=0$ and $\delta=\pi$, of the $\beta$-Beam based on $^{18}$Ne and $^6$He with standard fluxes and $L=130$ km (solid black line) to that achievable if the $^{18}$Ne flux is a factor 2 or 5 times smaller (dashed red and dotted blue lines respectively). In order to compensate the lower $^{18}$Ne flux, a larger $^6$He flux, twice the standard, was assumed and a running time of 2 years with $^6$He and 8 years with $^{18}$Ne was considered, instead of the 5+5 years of the standard scenario. 
The expected efficiencies and beam-induced backgrounds of the detector when exposed to the considered beams have been added as migration matrices extracted from Ref.~\cite{Burguet-Castell:2005pa}. In all the simulations the following best fit values and $1 \sigma$ errors for the known oscillation parameters were assumed $\Delta m^2_{21} = (7.7 \pm 0.3)\cdot 10^{-5}$ eV$^2$, $\Delta m^2_{31} = (2.4 \pm 0.1)\cdot 10^{-3}$ eV$^2$, $\theta_{12} = 34.0 \pm 1.3$ and $\theta_{23} = 45.0 \pm 4.5$. These parameters were marginalized over to present the final curves. The background caused by atmospheric neutrinos in the detector was neglected in this section but will be discussed in detail in the next section. The evaluation of the performance of the facility made use of the GLoBES software \cite{Huber:2004ka,Huber:2007ji}. 

As can be seen in Fig.~\ref{fig:lowne}, in the more optimistic case in which the $^{18}$Ne flux is only a factor 2 smaller than the standard, the sensitivity is not seriously affected provided that a factor 2 larger flux of $^6$He is achievable. However, as we will discuss in the next section, the longer time required running with a low $^{18}$Ne signal, translates in a less favorable signal/background fraction for the atmospheric neutrino background and more demanding suppression factors would be needed compared to the standard fluxes in order to achieve a similar performance. For a factor 5 reduction in the $^{18}$Ne flux the sensitivity gets more degraded.  

A similar analysis of the achievable fluxes for $^8$Li and $^8$B is only in its very early stages. Assuming that the two ions can be produced with the rates described in \cite{Rubbia:2006pi}, preliminary estimates of their fluxes after the acceleration up to $\gamma=100$ through the SPS yield fluxes usually larger than the standard fluxes considered for their $^6$He and $^{18}$Ne counterparts. However, there are still many uncertainties in these estimations, we will therefore present our result for three different assumptions of the achievable $^8$Li and $^8$B fluxes. We will consider for definiteness the  same ``standard'' fluxes of $1.1 \cdot 10^{18}$ decays per year for $^8$B and $2.9 \cdot 10^{18}$ decays per year for $^8$Li as are usually considered for $^{18}$Ne and $^6$He and study the effect of scaling those numbers by factors of 2 and 5.

\begin{figure}[t!]
\vspace{-0.5cm}
\begin{center}
\begin{tabular}{cc}
\hspace{-0.55cm} \epsfxsize7.5cm\epsffile{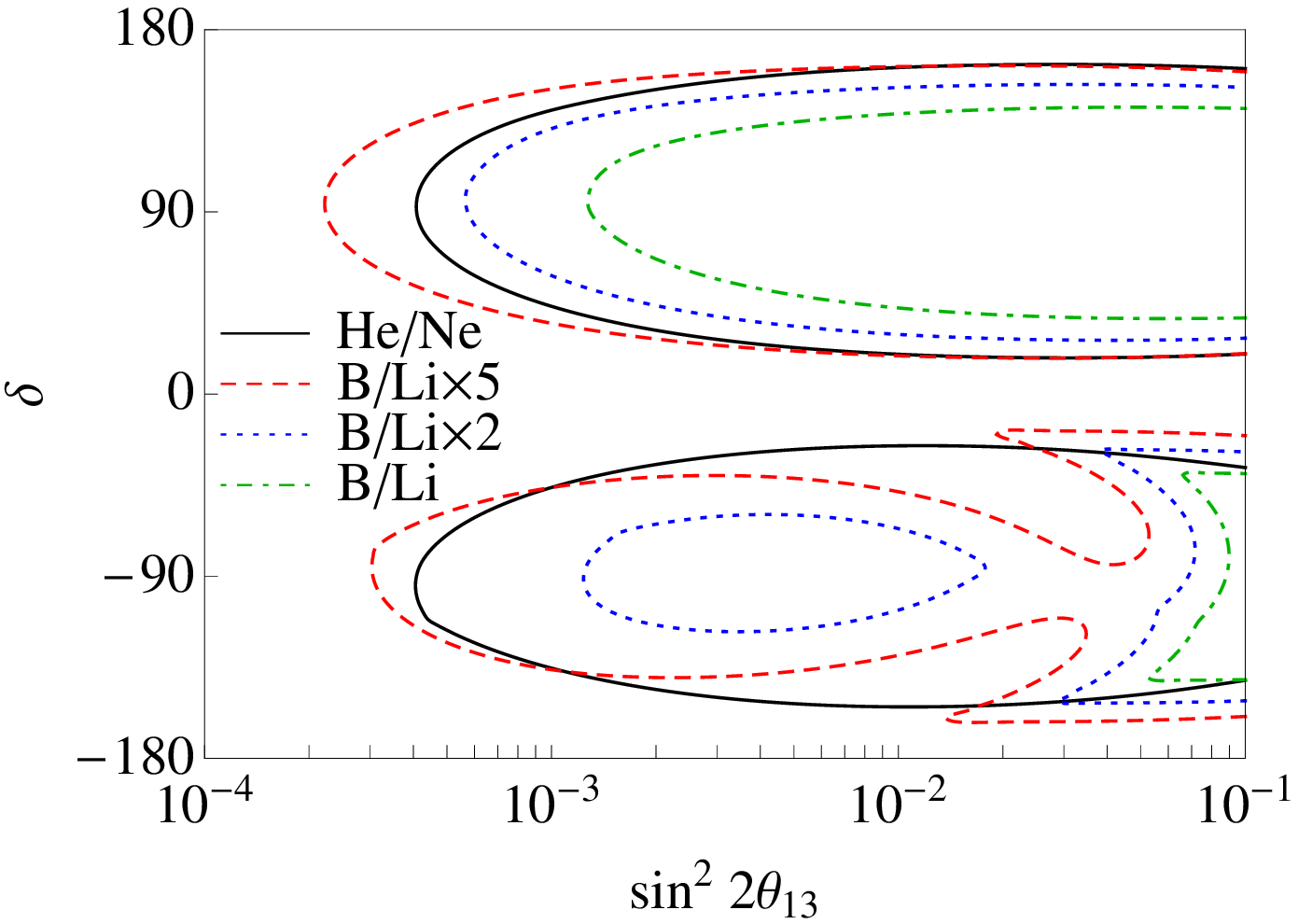} & 
                 \epsfxsize7.5cm\epsffile{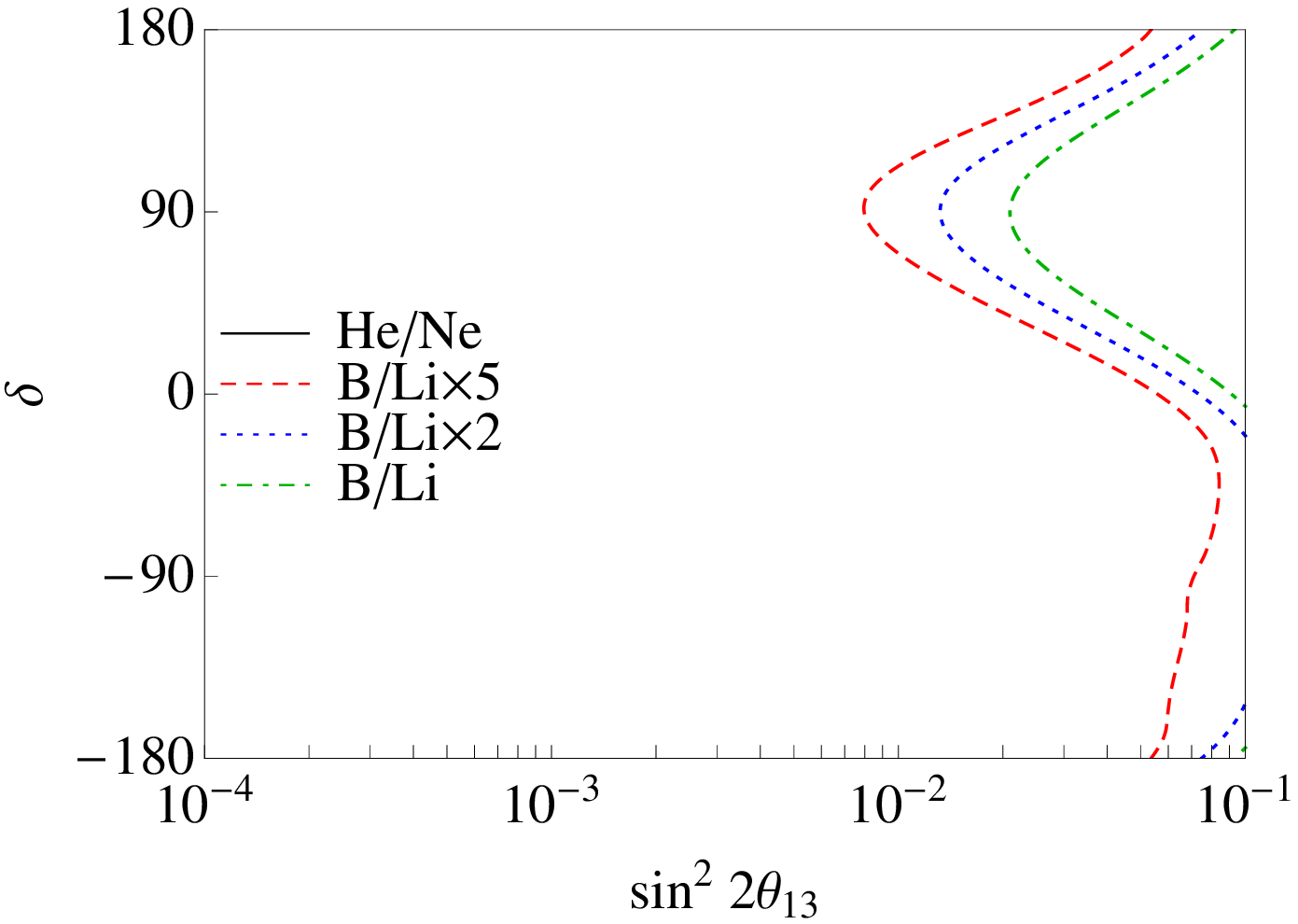} \\
\hspace{-0.55cm} \epsfxsize7.5cm\epsffile{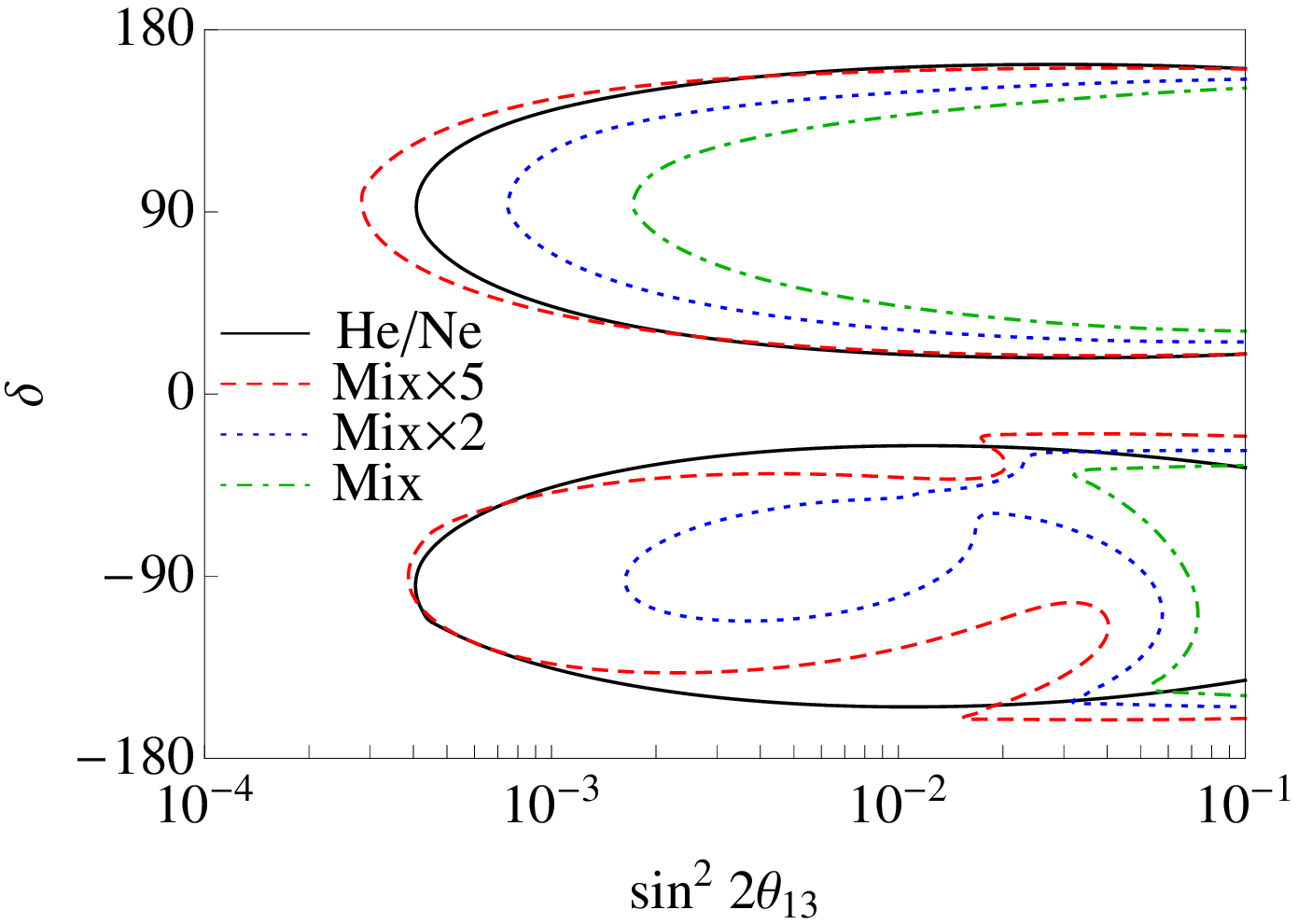} & 
                 \epsfxsize7.5cm\epsffile{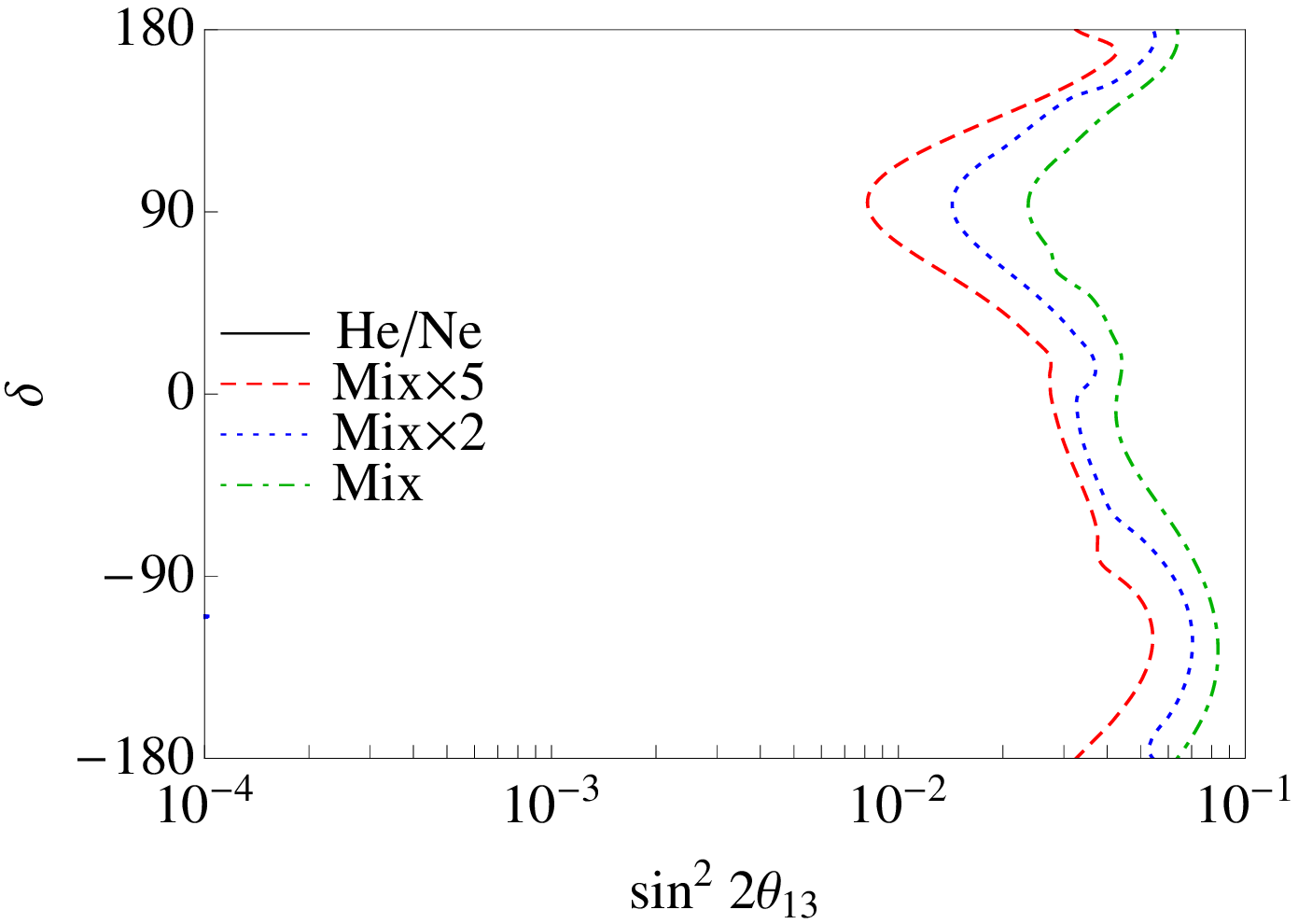} 
\end{tabular}
\caption{\label{fig:setup}
Comparison of the discovery potential to CP violation (left panels) and to a normal mass hierarchy (right panels) of the standard setup (black solid lines) and setups 2 (upper panels) and 3 (lower panels) defined in the text.}
\end{center}
\end{figure}

With the four ion candidates and two baselines available we have explored the sensitivities to the unknown parameters of three possible $\beta$-Beam setups:
\begin{itemize}
\item Setup 1. The ``standard'' setup with neutrinos from $^{18}$Ne and $^6$He decays accelerated to $\gamma=100$. The baseline is $L=130$ km and the fluxes are the standard $1.1 \cdot 10^{18}$ and $2.9 \cdot 10^{18}$ decays per year for $^{18}$Ne and $^6$He respectively. We assumed a 5 year run with each ion. This setup is represented by the black solid line in Fig~\ref{fig:setup}.

\item Setup 2. A setup based in $^8$Li and $^8$B decays accelerated to $\gamma=100$. The baseline is $L=650$ km and the fluxes are 1, 2 and 5 times the standard.
 We assumed a 5 year run with each ion. This setup is represented in the upper panels of Fig~\ref{fig:setup}.

\item Setup 3. A setup based in $^8$Li, $^8$B and $^6$He decays accelerated to $\gamma=100$. The baseline is $L=650$ km the same fluxes as for setups 1 and 2 were assumed. We assumed a 5, 3 and 2 year run with $^8$B, $^8$Li and $^6$He respectively. This setup is represented in the lower panels of Fig~\ref{fig:setup}.
\end{itemize}

The expected CP discovery potential, defined as the values of $\theta_{13}$ and $\delta$ that would allow to discard at $3 \sigma$ the CP conserving values $\delta=0$ and $\delta=\pi$, is presented in the left panels of Fig~\ref{fig:setup}. Setup 2 (upper panel) and setup 3 (lower panel) are compared to the standard setup (solid black lines). The discovery potential down to the smallest values of $\theta_{13}$ corresponds to setup 1, since the shorter baseline guarantees higher statistics. Setups 2 and 3 require a flux 5 times larger in order to reach smaller values of $\theta_{13}$ than setup 1. Moreover, the stronger matter effects present at these higher energies and baseline can mimic true CP violation and lead to degeneracies~\cite{Minakata:2001qm} that translate into the loss of sensitivity for negative values of $\delta$ around $\sin^2 2 \theta_{13} = 3 \cdot 10^{-2}$. In order to alleviate this degeneracy problem we introduced setup 3, the combination of information at the first oscillation peak from $^8$B and $^8$Li decays with that from $^6$He decays at the second oscillation peak can alleviate the sign degeneracies and help to fill in the gap in sensitivity for negative values of $\delta$ \cite{Donini:2006dx}. If the flux of $^8$B and $^8$Li is high enough, the degeneracies are already much reduced by the larger statistics and the gain provided by setup 3 is smaller.

The term in the oscillation probability that provides the sensitivity to CP violation is suppressed by $\sin 2 \theta_{13}$ and $\Delta m^2_{21} L/E$ and has to compete with a $\delta$-independent term suppressed by $\sin^2 2 \theta_{13}$. The sensitivity to CP violation thus decreases for the largest values of $\theta_{13}$, where the $\delta$-independent term dominates. However, in setups 2 and 3 the value of $L/E$ is larger for the lower energy bins than in setup 1 and thus, $\Delta m^2_{21} L/E$ being larger, their sensitivity to CP violation can outperform that of setup 1 for large values of $\theta_{13}$ even if their statistics is lower. For instance, with only a factor 2 increase in the flux, the $^8$Li and $^8$B setup is already outperforming setup 1 for negative values of $\delta$ and large $\theta_{13}$ and for a factor 5 increase in the flux the performance of setup 2 is always better than setup 1, except in the degeneracy region, even though the factor 5 does not fully compensate the $L^{-2}$ suppression from  the longer baseline. 

The expected discovery potential to a normal mass hierarchy, defined as the values of $\theta_{13}$ and $\delta$ that would allow to discard an inverted hierarchy at $3 \sigma$, is presented in the right panels of Fig~\ref{fig:setup}.  Setup 2 (upper panel) and setup 3 (lower panel) are presented. Notice that only setups 2 and 3 show some sensitivity to mass hierarchy. Setup 1 has too small matter effects due to the low energy of the beam and the shorter baseline to be able to measure the mass hierarchy by itself. Nevertheless, some sensitivity to the mass hierarchy could be gained when combining its information with atmospheric neutrino oscillations measured at the detector \cite{Campagne:2006yx}. The better sensitivity of setup 3 with respect to setup 2 for negative values of $\delta$ is again due to the complementarity of the information on the oscillation probability of $^6$He at the second oscillation peak that allows to break degeneracies when combined with $^8$Li and $^8$B at the first peak \cite{Donini:2006dx}. 

We conclude that the setups 2 and 3 based on $^8$Li and $^8$B decays can outperform the standard setup 1 based on $^{18}$Ne and $^6$He decays if fluxes larger than at least 2 times the standard can be achieved. For a flux 5 times the standard the setups based in $^8$Li and $^8$B beams outperform the standard setup for all the parameter space except the area around $\sin^2 2 \theta_{13} = 3 \cdot 10^{-2}$ and negative $\delta$ where degeneracies spoil their performance. For fluxes not much larger than 2 times the standard, setup 1 will perform better for smaller values of $\theta_{13}$, while setups 2 and 3 can provide sensitivity to the mass hierarchy and also cover a larger fraction of parameter space for CP discovery potential if $\sin^2 2 \theta_{13}>3-5 \cdot 10^{-2}$. These setups can then be a very interesting alternative to probe leptonic CP violation and the mass hierarchy if the present hint for large $\theta_{13}$ is confirmed by the next generation of reactor and accelerator experiments. Indeed in Refs.~\cite{Winter:2008cn,Winter:2008dj} it was shown that, if a factor 5 in luminosity is achieved, a setup based on $^8$Li and $^8$B decays accelerated to $\gamma=100$ can confirm at $5 \sigma$ the $\theta_{13}$ measurement and discover at $3 \sigma$ the mass hierarchy for any value of $\delta$ and CP violation for $80 \%$ of the values of $\delta$ if $\theta_{13}$ is large enough to be seen by Double Chooz.  

\section{The atmospheric neutrino background}
\label{atmo}

\begin{table}
\begin{center}
\begin{tabular}{|c|c|c|c|c|} \hline \hline
    & $^{18}$Ne & $^{6}$He & $^{8}$B & $^{8}$Li \\ 
\hline
 Signal ($\delta=90^\circ$) &   446 &     265      &       86         &     37    \\
~ Signal ($\delta=-90^\circ$) ~ &    189   &      651         &     37        &     104   \\
 Beam back.           &   51    &      37         &       11       &    12    \\ 
 Atmo. back.   &  ~ 11144~  &   ~  11144 ~    &   ~ 814   ~     &   ~ 814  ~  \\ 
\hline
\hline
\end{tabular}
\caption{\label{tab:rates} Event rates for the signal for $\theta_{13}=5^\circ$ and the background of the four ions considered at $L=130$ km and $L=650$ km for He/Ne and B/Li respectively.
Efficiencies have been included in the computation of the rates. The expected atmospheric background rate after the angular cut but before the timing suppression factor is also shown.}
\end{center}
\end{table}
One of the main sources of background that can spoil the $\beta$-Beam sensitivity is the background from atmospheric neutrinos. This background can be reduced by imposing angular cuts in the direction of the beam. Indeed, even if the direction of the incoming neutrino cannot be measured, it is increasingly correlated with the direction of the detected muon at higher energies. This situation is depicted in Fig.~5 of Ref.~\cite{Burguet-Castell:2005pa} for neutrinos from $^{18}$Ne and $^6$He with $\gamma$ factors of $120$, $150$ and $350$. In particular the energy spectrum of of neutrinos from $^8$Li and $^8$B decays with $\gamma=100$ is very similar to $^{18}$Ne and $^6$He at $\gamma=350$ since their decay energy is precisely $\sim 3.5$ times larger. As can be seen from the figure, the mean angle between the muon and the incoming neutrino is much smaller in the $\gamma = 350$ scenario. In fact, an angular cut requiring $90 \%$ of the efficiency was applied in \cite{Burguet-Castell:2005pa} and, in mean, was found to correspond to  $\cos \theta_l > -0.5$ for $\gamma = 120$ and  $\cos \theta_l > 0.45$ for $\gamma = 350$. The cut is included in the migration matrices extracted from Ref.~\cite{Burguet-Castell:2005pa} and used here. This angular cut translates in a reduction of the atmospheric background that is three times stronger for the higher energy scenario.

In order to estimate the atmospheric neutrino background arriving in the same direction of the beam, we have evaluated the expected number of oscillated atmospheric muon neutrinos using the flux from \cite{honda} for the Frejus site that arrived within a solid angle $\sim \sqrt{1/E\textrm{(GeV)}}$ from the beam direction for the different energy bins considered and studied its impact on the sensitivities. As can be seen from the event rates reported in Tab.~\ref{tab:rates}, even after imposing this directional cut, the background level dominates the expected signal and an additional cut must be imposed to reduce it to acceptable levels. This can be achieved by accumulating the signal in small bunches so as to use timing information to reduce the constant atmospheric background. If the storage ring has a total length $L$ and there are $N$ ion bunches circulating in it with a longitudinal spread $l$, when the ions decay in the straight sections of the beam traveling close to the speed of light, they will produce a pulsed neutrino signal. The time length of these pulses will just be $l/c$, while the periodicity with which a given ion bunch produces a pulse is $L/c$, timing this signal in the detector can therefore reduce the atmospheric background by a suppression factor given by: 
\begin{equation}
SF = \frac{Nl}{L},
\end{equation} 
since only that fraction of time corresponds to signal pulses. Previous analysis showed that, for the standard setup, the decaying ions must be accumulated in very small bunches so as to achieve a $SF = 10^{-2}-10^{-3}$ suppression factor of the background \cite{Burguet-Castell:2005pa,Mezzetto:2005ae}. 

\begin{figure}[t!]
\vspace{-0.5cm}
\begin{center}
\begin{tabular}{cc}
\hspace{-0.55cm} \epsfxsize7.5cm\epsffile{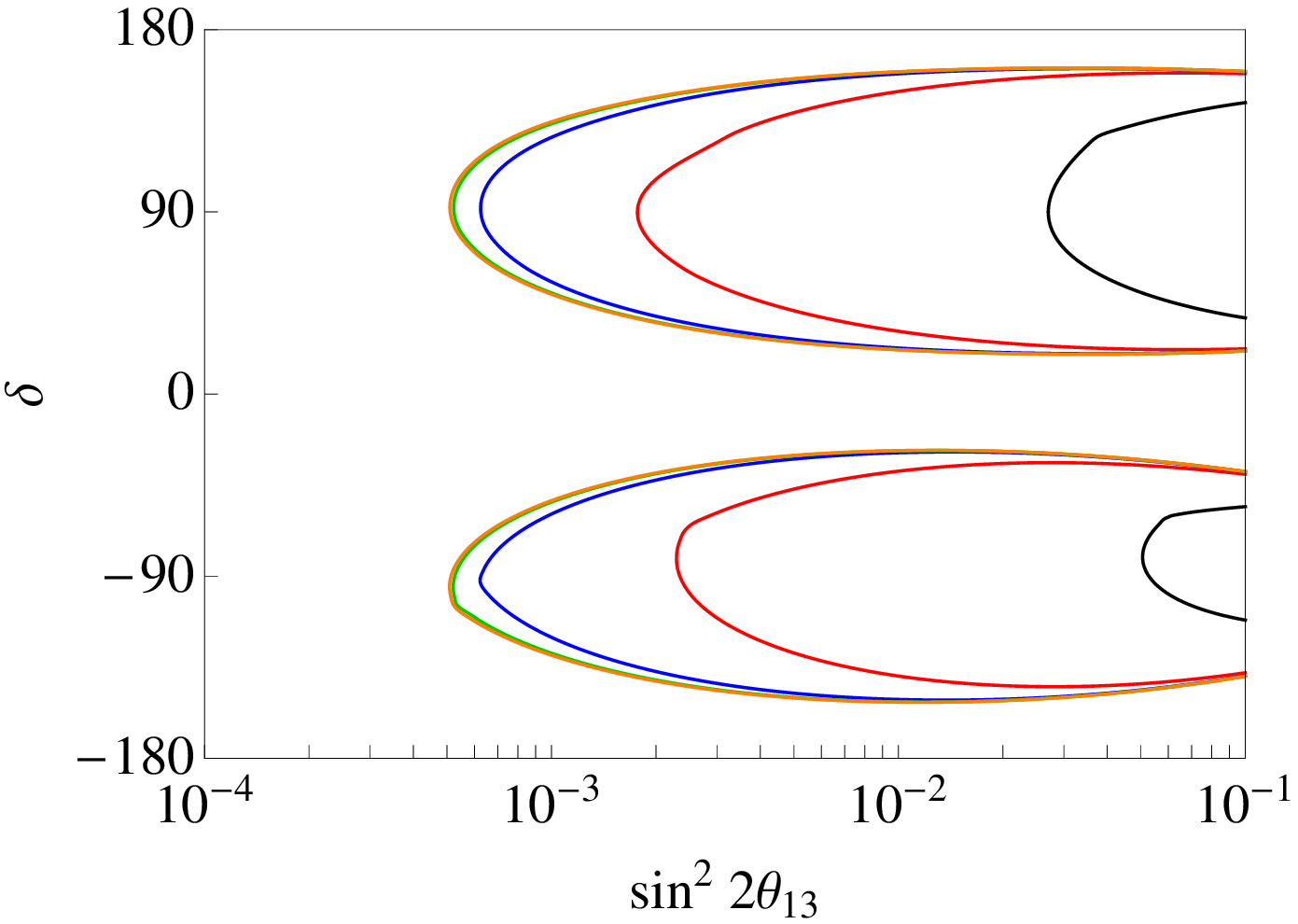} & 
                 \epsfxsize7.5cm\epsffile{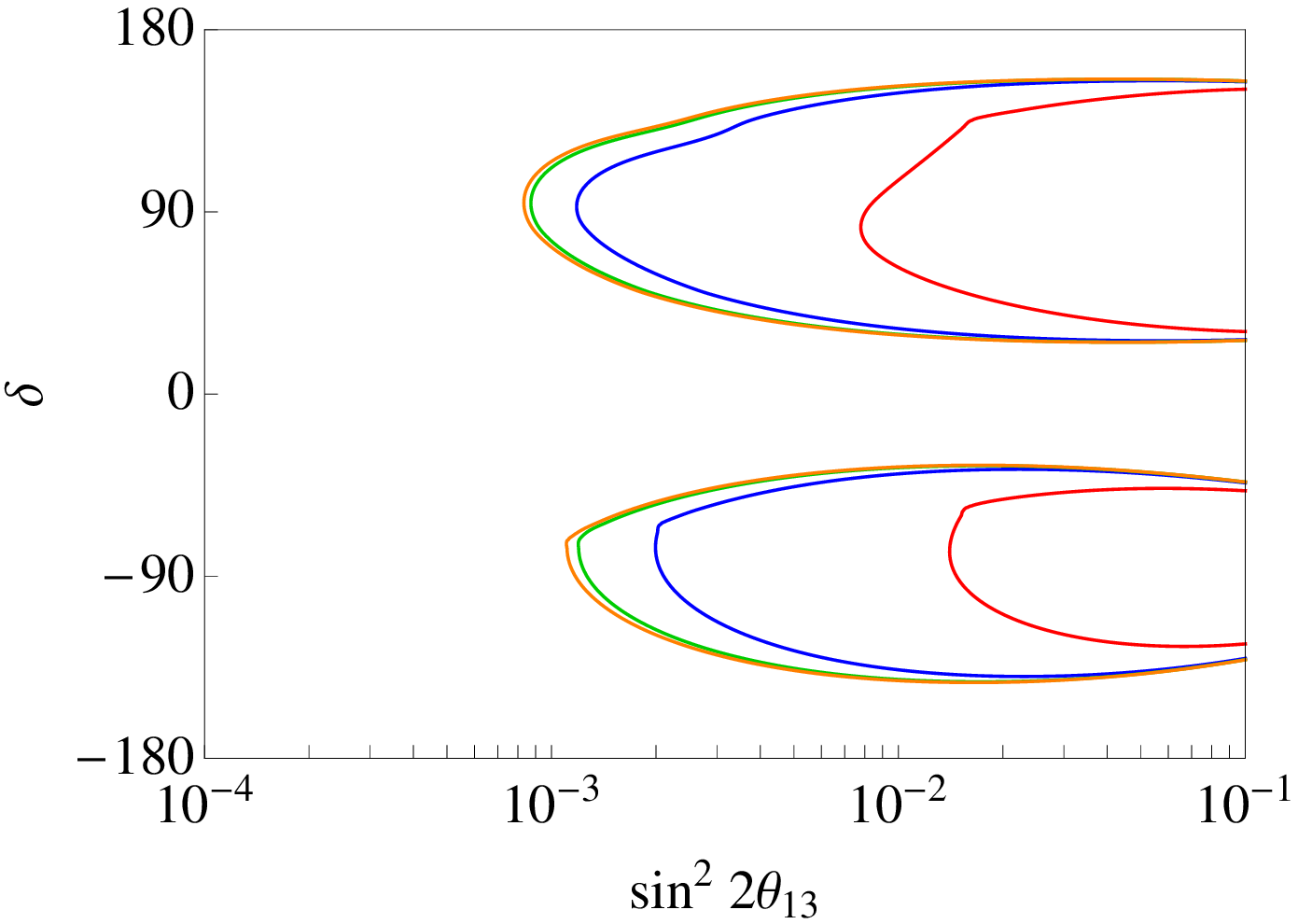} 
\end{tabular}
\caption{\label{fig:atmoHeNe}
Dependence on the background suppression factor of the the 3 $\sigma$ discovery potential to leptonic CP violation of the standard $\beta$-Beam scenario with 2 times larger flux of $^6$He but 2 (5) times smaller $^{18}$Ne flux in the left (right) panel.
The lines correspond to suppression factors of $0$, $10^{-4}$, $10^{-3}$, $10^{-2}$ and $10^{-1}$.}
\end{center}
\end{figure}

In Fig.~\ref{fig:atmoHeNe} we show the dependence on the achievable SF of the discovery potential to CP violation of the standard scenario with a reduced $^{18}$Ne flux. As in the previous section we assumed that a flux twice the standard is achievable for $^6$He, i.e. $5.8 \cdot 10^{18}$ decays per year, while the flux of $^{18}$Ne is reduced two or five times corresponding to $5.5 \cdot 10^{17}$ and $2.2 \cdot 10^{17}$ decays per year in the left and right panels respectively. An asymmetric running time of 2 years with $^6$He and 8 with $^{18}$Ne was assumed. The lines correspond to suppression factors of $0$ (orange), $10^{-4}$ (green), $10^{-3}$ (blue), $10^{-2}$ (red) and $10^{-1}$ (black). As can be seen a suppression factor of $10^{-4}$ is practically identical to the case in which the background was completely neglected. As for the standard case, in the scenario with a factor 2 suppression of the $^{18}$Ne (left panel) a suppression factor of the background by $10^{-3}$ does not spoil much the sensitivity with respect to the no-background limit and can then be viewed as a goal to achieve the best performance from the facility. If $\sin^2 2\theta_{13} > 10^{-2}$ and the present hint is confirmed by the next generation of facilities, suppression factors as big as $10^{-2}$ are acceptable without seriously spoiling the CP discovery potential. However, in the case the flux of $^{18}$Ne is a factor 5 smaller than the standard flux, a stronger suppression of the atmospheric background would be necessary to avoid too much sensitivity loss. 

Since the atmospheric neutrino background decreases with the energy as $\sim E^{-2}$, for the other setups considered here based on $^8$Li and $^8$B decays with $\sim 3.5$ times higher energy, an order of magnitude less atmospheric background can be expected at the detector. We studied whether this allows to relax the stringent bunching required on the signal. Notice that the achievable ion flux is strongly affected by this strong requirement and a relaxation of this value could allow an increase in statistics.

\begin{figure}[t!]
\vspace{-0.5cm}
\begin{center}
\begin{tabular}{cc}
\hspace{-0.55cm} \epsfxsize7.5cm\epsffile{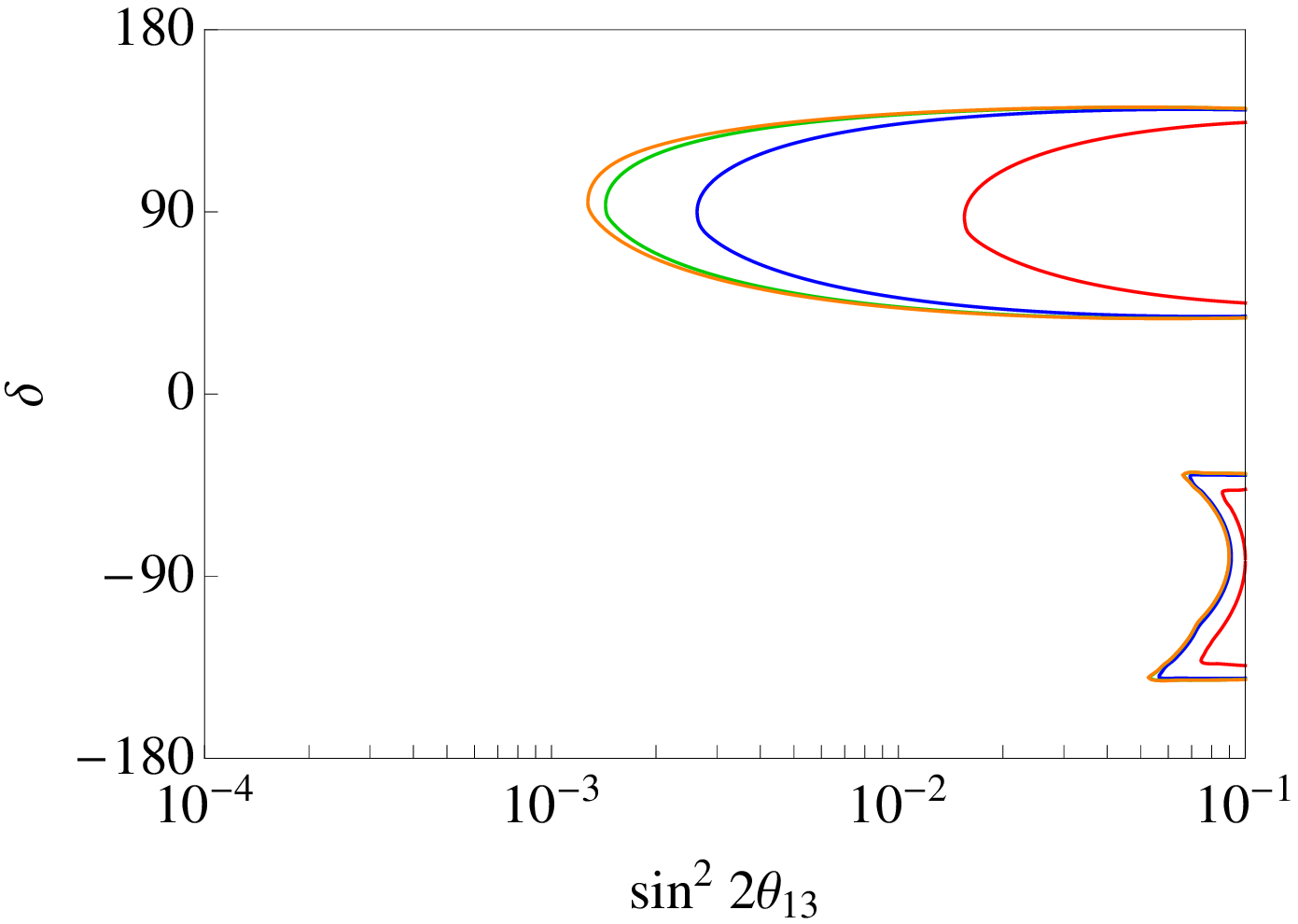} & 
                 \epsfxsize7.5cm\epsffile{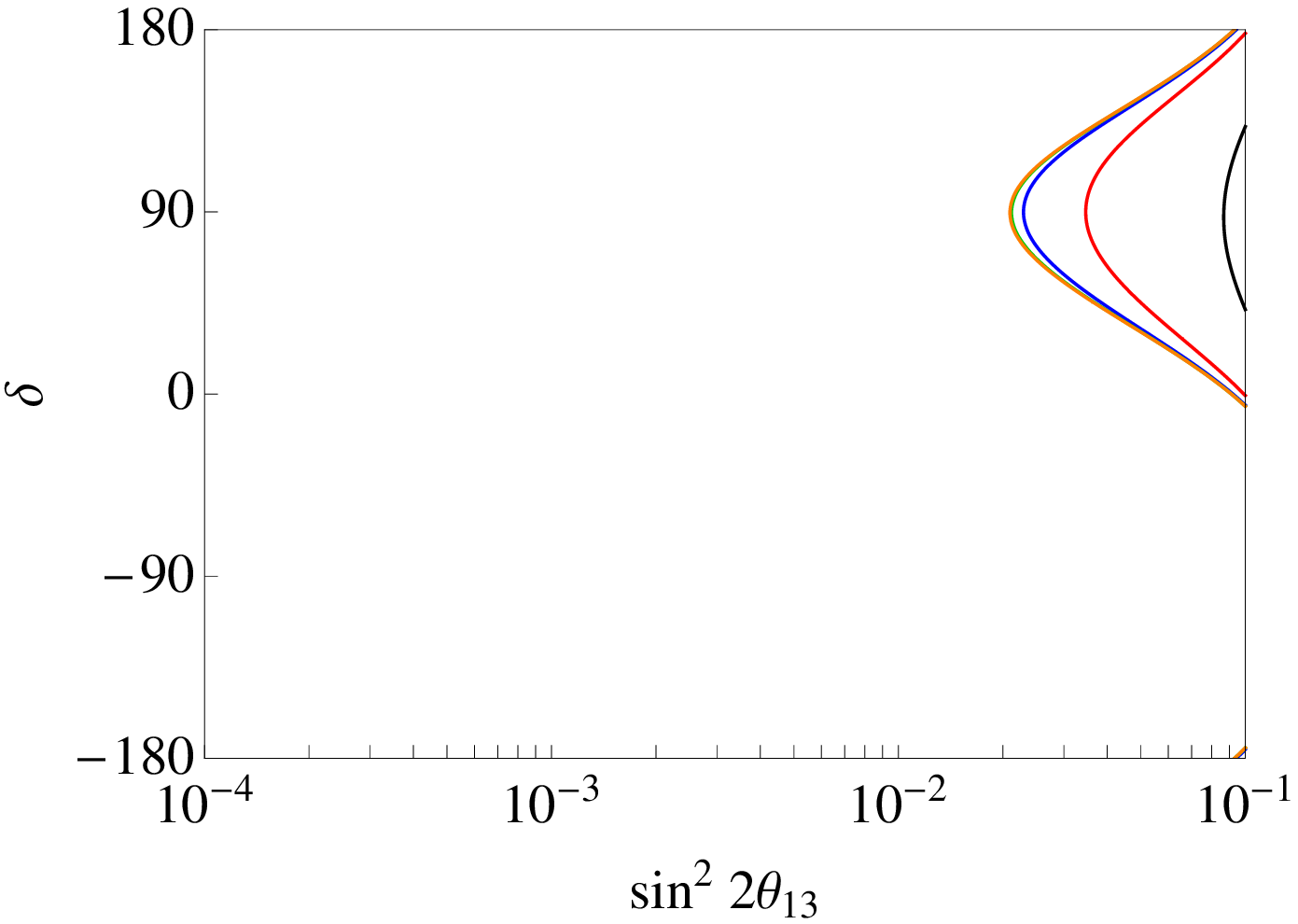} \\
\hspace{-0.55cm} \epsfxsize7.5cm\epsffile{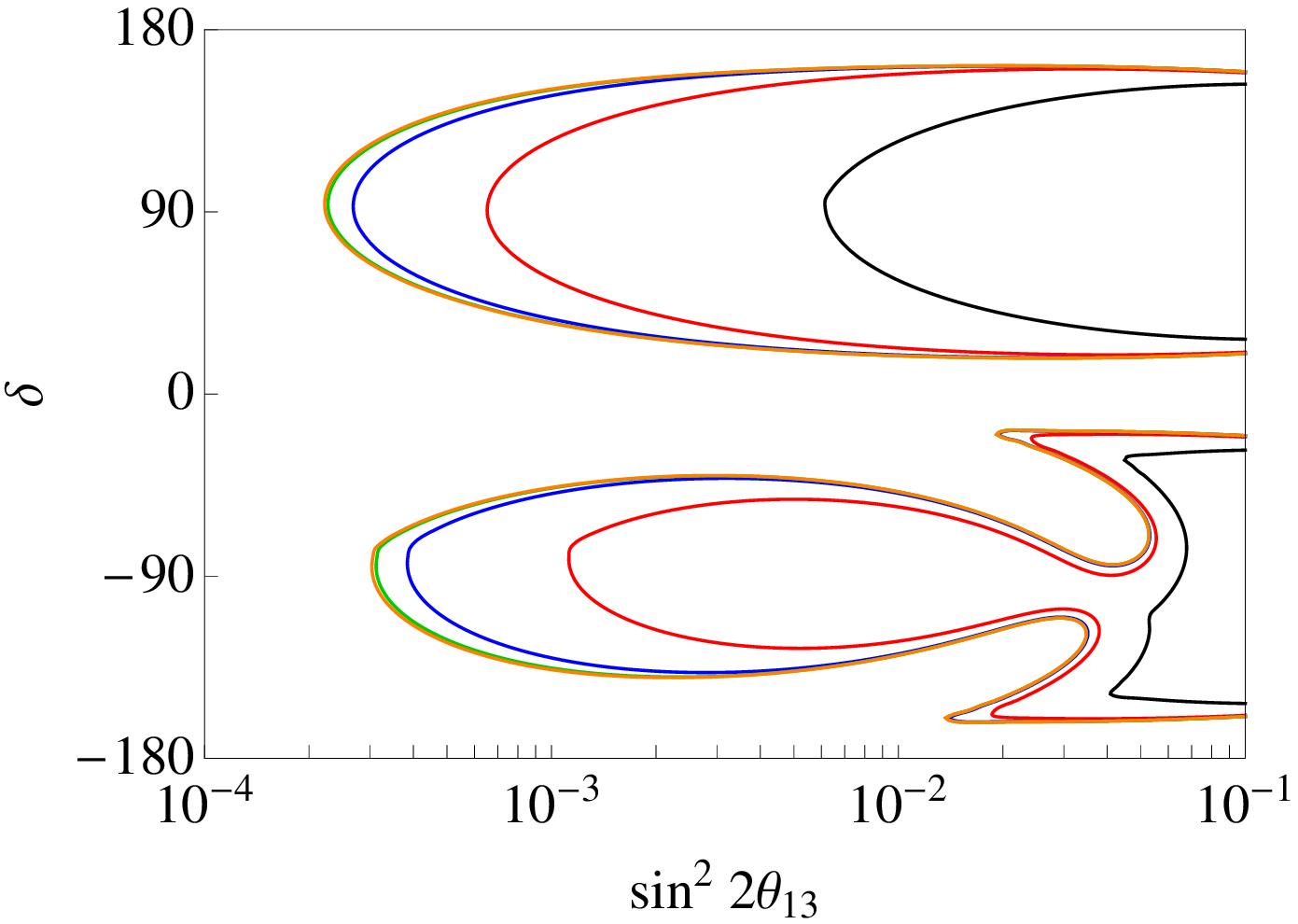} & 
                 \epsfxsize7.5cm\epsffile{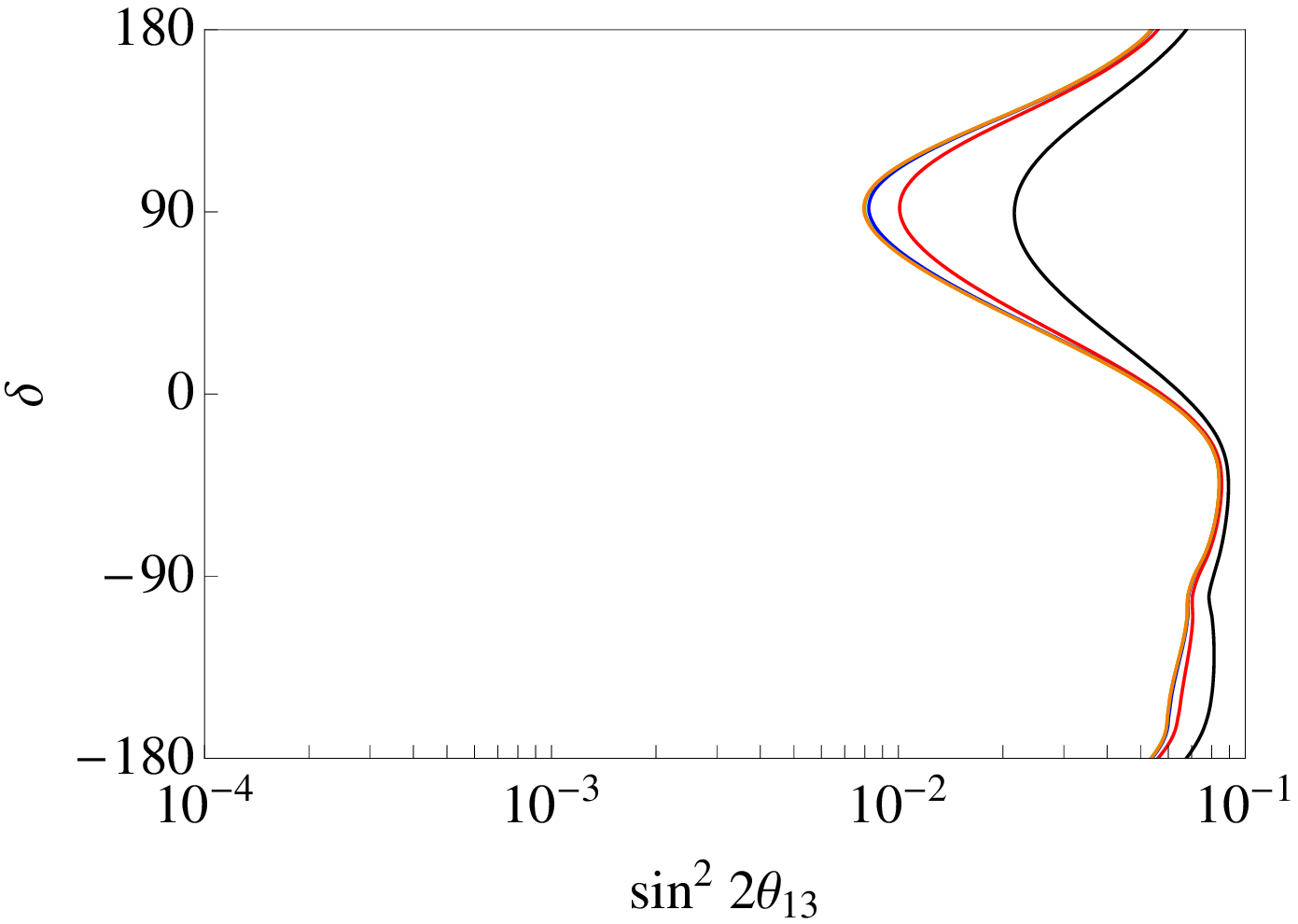} 
\end{tabular}
\caption{\label{fig:atmo}
Dependence on the background suppression factor of the the 3 $\sigma$ discovery potential to leptonic CP violation (left panels) and the mass hierarchy (right panels) of the $^8$B and $^8$Li setup. The lines correspond to suppression factors of $0$, $10^{-4}$, $10^{-3}$, $10^{-2}$ and $10^{-1}$. The upper panels correspond to fluxes of $^8$B and $^8$Li of $1.1 \cdot 10^{18}$ and $2.9 \cdot 10^{18}$ decays per year respectively and the lower panels to a flux increase by a factor 5.}
\end{center}
\end{figure}

In Fig.~\ref{fig:atmo} we show the dependence on the background suppression factor of the the 3$\sigma$ discovery potential to leptonic CP violation (left panels) and the mass hierarchy (right panels) of the $^8$B and $^8$Li setup. The upper panels correspond to the most pessimistic flux of $2.9 \cdot 10^{18}$ and $1.1 \cdot 10^{18}$ decays per year for $^8$Li and  $^8$B respectively. The lower panels correspond to the most optimistic assumption with a factor 5 larger fluxes. Notice that, as for the lower energy case, there is almost no difference between the sensitivity with no atmospheric background (orange line) and with a $10^{-4}$ background (green line). The atmospheric background thus becomes negligible if a $10^{-4}$ suppression is achieved. The sensitivities for the $10^{-3}$ suppression case (blue lines) are only slightly worse than those for $10^{-4}$ in the high statistics case. Thus, $10^{-3}$ or even $5 \cdot 10^{-2}$ seems the goal that should be achieved in order to exploit the full potential of the  $\beta$-Beam for small values of $\theta_{13}$ if the higher fluxes can be achieved. For $\sin^2 2 \theta_{13} > 10^{-2}$ suppression factors as large as $10^{-2}$ do not imply a sensitivity loss with respect to the case of no atmospheric background. On the other hand, in the low flux scenario, the smaller signal also demands a further reduction of the background. 

We conclude that, for the fluxes assumed, a background suppression around $\sim 10^{-3}-10^{-2}$ is required to achieve the full sensitivity of the $\beta$-Beam, depending on the size of $\theta_{13}$. The fact that the higher energy beams provided by $^8$Li and $^8$B decays require a similar background suppression than those based on $^{18}$Ne and $^6$He can be easily understood. On the one hand, the expected background is more than an order of magnitude lower for the $^8$Li and $^8$B setups, since the atmospheric background decreases with $E^{-2}$ and the energy is $\sim 3.5$ larger and the angular cut performed of $\sim \sqrt{1/E\textrm{(GeV)}}$ is more efficient at higher energies. On the other hand, the signal decreases due to the longer baseline with a factor $(650/130)^2=25$ which is not fully compensated by the larger cross section at higher energies (more than a factor $3.5$ larger since the cross section grows faster than linearly at those energies). Moreover, the efficiency of the detector gets degraded at high energies, since water Cerenkov detectors are not optimal beyond the quasielastic regime of the cross section. In particular, following Ref.~\cite{Burguet-Castell:2005pa}, we find the efficiencies to be more than a factor two smaller for the higher $Q$ ions than for $^{18}$Ne and $^6$He. This means that the signal/background fraction remains similar in all scenarios and similar suppression factors are required to achieve the full $\beta$-Beam potential. In the case where the flux of $^8$B and $^8$Li was assumed to be 5 times larger, the suppression factor could be somewhat relaxed beyond $10^{-3}$, given the larger signal, or even beyond $10^{-2}$ for large values of $\theta_{13}$. On the other hand, for the most pessimistic assumptions for the fluxes of $^8$B and $^8$Li and in the scenario with a factor 5 smaller $^{18}$Ne, even more stringent suppression factors, smaller than $10^{-3}$, would be desirable.

\section{Conclusions}
\label{concl}

We have compared the performance of several $\gamma=100$ $\beta$-Beams based on the decay of different ions. In particular we have studied setups exploiting the decay of $^8$Li and $^8$B at a baseline of $650$ km, which provide the complementary $L/E$ information of a wider neutrino spectrum and stronger matter effects, at the cost of lower statistics given the longer baseline required for neutrino oscillations to develop. 

We found that the setups involving $^8$Li and $^8$B always outperform the standard setup with $^{18}$Ne and $^6$He decays in sensitivity to the mass hierarchy, given the stronger matter effects provided by the higher energy and longer baseline. Moreover, the higher values of $L/E$ accessible at the lower end of the spectrum enhance the CP violation signal for large values of $\theta_{13}$ and can provide the $^8$Li and $^8$B setups with better coverage of the $\delta$ parameter space for the CP discovery potential. On the other hand, the reduced statistics due to the longer baseline required for the oscillation of $^8$Li and $^8$B neutrinos imply that the standard setup is preferable for small values of $\theta_{13}$, provided that the present pessimistic estimation for the $^{18}$Ne flux can be improved. Otherwise, the sensitivity of the standard setup is also deteriorated and $^8$Li and $^8$B setups might be preferable, depending on their achievable fluxes. Conversely, the stronger matter effects present for the $^8$Li and $^8$B setups, can mimic true CP violation and sign degeneracies difficult the discovery of CP violation at negative values of $\delta$. This situation could be improved by adding $^6$He antineutrinos at the same $650$ km baseline that would be at the second peak of its oscillation with the sign degeneracies at different places in the parameter space~\cite{Donini:2006dx}. The information from atmospheric neutrino oscillations at the detector could also provide similar sensitivities to the mass hierarchy when combined with the $\beta$-Beam data and improve the CP discovery potential~\cite{Campagne:2006yx,Donini:2007qt}.

Finally, we also studied the impact of the atmospheric background in the sensitivities and the suppression factor to be achieved through the timing of the ion beam in the storage ring in order to reduce it to acceptable levels. We found that to achieve the full performance of the $\beta$-Beam, suppression factors of $\sim 10^{-3}$ are desirable for all the setups studied. This suppression factor can be relaxed to $\sim 10^{-2}$ if $\sin^2 2 \theta_{13} > 10^{-2}$, that is, if the present hint for large $\theta_{13}$ is confirmed by the next generation of accelerator and reactor experiments.

To summarize, we believe that $\gamma=100$ $\beta$-Beam setups based on $^8$Li and $^8$B decays are a very interesting alternative to the standard setup, specially for large values of $\theta_{13}$. However, larger fluxes than the standard for $^{18}$Ne and $^6$He are required to achieve competitive CP discovery potential. We also found that matter effects induce sign degeneracies that spoil the sensitivity to the mass hierarchy and the discovery of CP violation for negative values of $\delta$. These degeneracies can be alleviated either by an increase of statistics or by the complementary information provided by $^6$He antineutrinos at the second oscillation peak. 

\section*{Acknowledgments}

I want to thank A.~Donini, C.~Hansen, P.~Hern\'andez, M.~Mezzetto and E.~Wildner for interesting discussions and for encouraging me to perform this study.
I am particularly indebted to O.~Mena for her help with the simulation of the atmospheric neutrino background and for cross-checking my results.
I acknowledge support by the DFG cluster of excellence ``Origin
and Structure of the Universe'' and from the European Community under the European Commission Framework Programme
7 Design Study: EUROnu, Project Number 212372.

\end{document}